Itinerant metamagnetism of $CeRu_2Si_2$ : bringing out the dead.
Comparison with the new $Sr_3Ru_2O_7$ case

J. Flouquet, P. Haen*, S. Raymond, D. Aoki* and G. Knebel

CEA/DSM, Service de Physique Statistique, Magnétisme et Supraconductivité, 17 rue des Martyrs, 38054 Grenoble cedex, France
* CNRS/CRTBT, 25 rue des Martyrs, 38042 Grenoble cedex

**Abstract** : Focus is given on the macroscopic and microscopic experimental works realized during a decade on the clear case of itinerant metamagnetism in the heavy fermion paramagnetic compound $CeRu_2Si_2$ . Emphasis is made on the feedback between the band structure, the exchange coupling and the lattice instability. Sweeps in magnetic field, pressure and temperature feel the pseudogap of this strongly correlated electronic system as well as its equivalent $CeRu_2Ge_2$ at a fictitious negative pressure. Some mysteries persist as the complete observation of the FS above the metamagnetic field $H_M$ and the detection of the dynamical ferromagnetic fluctuation near $H_M$. The novelty of the bilayer ruthenate $Sr_3Ru_2O_7$ is discussed by comparison. Despite differences in spin and electronic dimensionality many common trends emerge.

The name of metamagnetism was introduced for antiferromagnetic (AF) materials where at low temperature for a critical value of the magnetic polarization i.e. of the magnetic field (H) the spin flips (1). It gives rise to a first order phase transition. It was extended to paramagnetic (Pa) systems where field reentrant ferromagnetism (F) may appear notably in itinerant magnetism (2). Finally, it was also used to describe the case of a clear crossover inside a persistent paramagnetic state between a low field paramagnetic (Pa) phase and an enhanced paramagnetic polarized (PP) phase ; the main feature is a strong positive curvature in the magnetization (M) with a marked inflection point at $H = H_M$.

In itinerant systems, it is always possible to reproduce a strong non linearity of M in H by an effective density of states $\rho(E)$ picture with a pseudogap near the Fermi level $E_F$ (ie minimum of density of states). The magnetic field by polarizing the system, will drive $\rho(E_F)$ to a maximum for a given value of $H_M$ ; for a symetrical case, the relative field shift of spin up and spin down density of states will reach in phase the optimal decoupling condition (3-4).

With extensive studies of the tetragonal heavy fermion compound $CeRu_2Si_2$ , the experimentalists succeed to realize a large variety of experiments which lead to an excellent microscopic understanding of the feedback mechanisms between electronic, magnetic and lattice instabilities (5). The combination of the growth of excellent crystals and of the possibility of easy sweep of different effects by moderate pressure (P < 10 GPa) and magnetic field ($H_M$ = 7.8 T) has permitted to reveal major effects.

They stress that, under magnetic field, the system may be driven through a magnetic instability but prevented by quantum magnetic fluctuation (6). As the situation of $CeRu_2Si_2$ is highly documented, it is an excellent reference for the understanding of new examples of metamagnetism recently discovered as in the bilayer ruthenate $Sr_3Ru_2O_7$ (7). Comparison will be also given with $CeRu_2Ge_2$ which at P = 0 has a ferromagnetic ground state but above P* ~2.5 GPa an antiferromagnetic one similar to that found in $CeRu_2Si_2$ family and characterized by $P_C$ ~8 GPa (8). The study of $CeRu_2Si_2$ has led to a large number of papers notably from Cambridge, Frankfurt, Geneva and Osaka. We will describe here mainly experiments performed recently in Grenoble.

### $CeRu_2Si_2$ : proximity to magnetic quantum critical point and metamagnetism

In the axial lattice of $CeRu_2Si_2$, inside a single impurity Kondo frame, the bare Ce ion has an anisotropic doublet S=1/2 crystal field ground state mainly formed by the pure $|\pm5/2\rangle$ component of its J = 5/2 angular momentum. The Kondo temperature $T_K$ is near 20 K ie less than the crystal field splitting $\Delta$ ~ 200 K but rather comparable to the intersite exchange coupling (9). This single site Kondo approach must be changed for a lattice view at low temperature as the magnetic and electronic coherence will govern all behaviors (T < $T_K$).

The heavy fermion tetragonal lattice $CeRu_2Si_2$ is closed to a so called quantum critical point (QCP) at zero pressure (P = 0). For the pure lattice $CeRu_2Si_2$, the quantum critical pressure $P_C$ below which the system will transit to a long range antiferromagnetic ordering corresponds to a negative pressure near few kbar . Tiny expansion of the volume (V) by alloying $CeRu_2Si_2$ with La (10) or Ge (11) induces a magnetic order respectively above $x_C$ of 8% or 5%. Its proximity to QCP is also demonstrated by the unusual large value of its electronic parameter $\Gamma_e$ ~ 200.

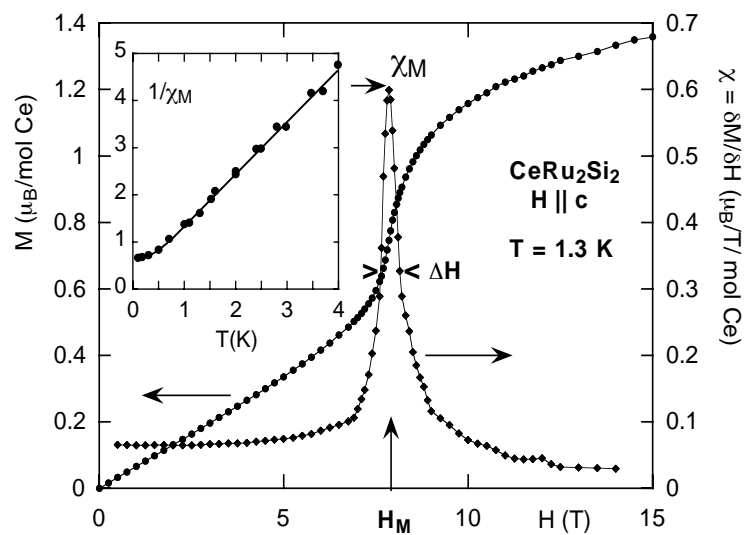

Figure 1 : Low temperature magnetization M (H) of $CeRu_2Si_2$, the insert is the temperature dependence of the differential susceptibility $\chi_M$ (see ref. 5)

By polarizing the lattice with a magnetic field (H), a strong non linear effect appears in the magnetization, M (H) ; a marked inflection point occurs for $H_M$ ~7.8 T (figure 1). While the change of macroscopic quantities is continuous through $H_M$, its location is well defined (5, 9). It corresponds to an enhancement of the average effective mass <m*> derived from thermodynamic measurements. It is also associated with huge magnetostriction and softening of the lattice. The differential susceptibility $\chi_M = \frac{\partial M}{\partial H}$ at $H_M$ on cooling has a Curie law dependence down to 1K; its flattening occurs only below 300 mK. Marked signatures appear in transport properties for example in the residual and inelastic contributions to the resistivity.(see 5 ).

The figure 2 represents the field variation of the coefficient $\gamma \sim <m^*>$ of the linear $\gamma T$ term of the specific heat (extrapolated at T = 0K) and of the field derivative of the magnetostriction V(H). An attempt has been even made (6) to scale in the vicinity of $H_M$ the field dependence of $\gamma$ by a $\sqrt{H - H_M}$ law with the idea that the magnetic field may play the role of a $\delta$ controlled parameter as concentration x or pressure to tune the system through AF QCP (12) ; deviations appear for $\frac{H - H_M}{H_M}$ lower than 0.3.

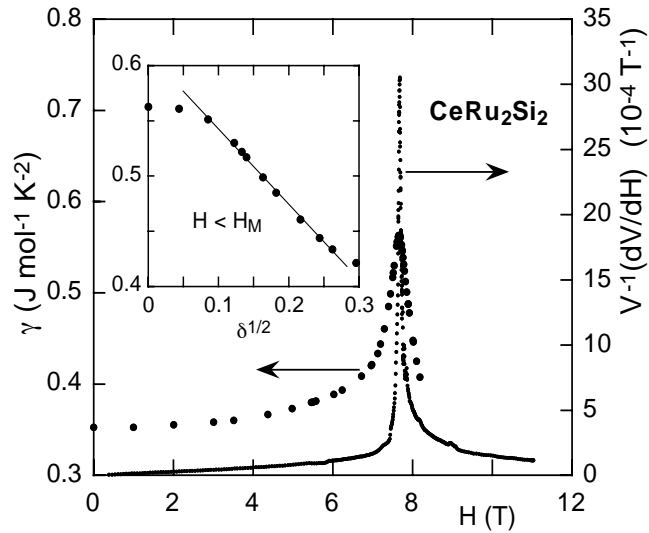

Figure 2 : Field variation of $\gamma = (C/T)_{T \to 0}$ and of the H derivative of the magnetostriction in $CeRu_2Si_2$ ; the insert shows $\gamma$ as a function of the square root of the tuned parameter

$$\Gamma = \frac{|H - H_M|}{H_M}$$ (see ref. 5 and 6)

As shown in figure 1 and 2, sharper field features occur in magnetostriction and differential susceptibility $\chi_M$ at $H_M$ measured at a constant ambient pressure ; the finite residual width ($\Delta H$) of the crossover regime at $H_M$ from $\chi_M$ is only 400 Oe ie $5.10^{-3}$ of $H_M$. At constant volume the transition though $H_M$ from $\chi_M$ is far more broadened than at constant pressure since the sharpness observed in $\chi_M$ (P = 0, T $\to$ 0) is driven by the huge magnetostriction contribution (13). The strong softening of the lattice (40%) suggest a field quasicollapse of this Kondo lattice (14-15-16) by analogy to the pressure collapse of the cerium metal (see 15).

NMR experiments on Ru nuclei (17) have clearly proved that there is no change in the hyperfine coupling through the metamagnetic transition. The H invariance in the microscopic nature of the magnetism is also confirmed in the spatial information given by the magnetization density map drawn by polarized neutron technic (18). The good scaling obtained in $1/T_1T$ variation of the $T_1$ relaxation time by the square of $\gamma$ verifies that $CeRu_2Si_2$ never crosses a magnetic instability at $H_M$ since each quantity should have a different T dependence at F or AF QCP (19). The emergence of a peak in $1/T_1T$ as well as in C/T in temperature for $H > H_M$ reflects also a main difference with the smooth continuous increase on cooling of those quantities below $H_M$. (20)

Microscopic measurements by neutron scattering show drastic variations in the nature of the magnetic correlations (21). The zero field incommensurate antiferromagnetic (AF) correlations are replaced by dominant ferromagnetic one above $H_M$ ; de Haas Van Alphen (dHVA) experiments prove clearly that this phenomena is associated with a deep modification of the Fermi surface (FS). A simple picture is that, below $H_M$ the 4f electron must be treated as itinerant and above $H_M$ as localized ie the FS is rather similar to the isoelectronic non magnetic compound $LaRu_2Si_2$ (22-23).

Well defined crossover field temperature $T_\alpha$ (H) phase diagram.

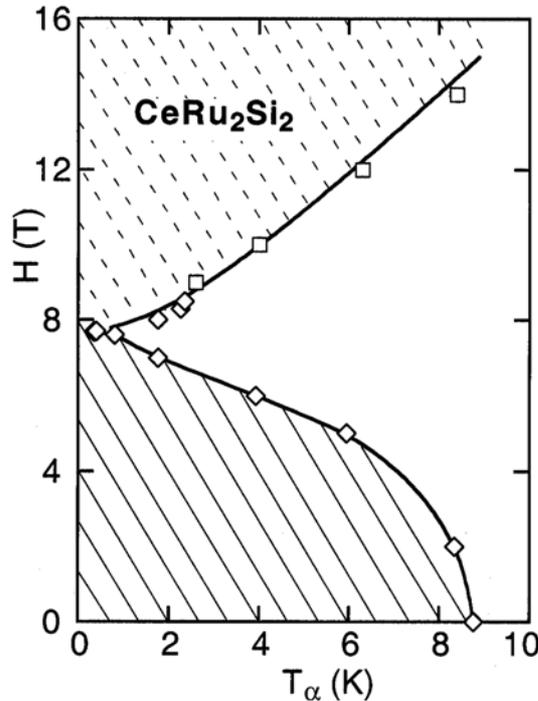

Figure 3 : The crossover phase pseudo-diagram $T\alpha$ (H) derived from thermal expansion measurements [◊]. The high field data [□] are the temperature of the C/T maxima observed in ref. [6].

Thermal expansion ($\alpha$) experiments at constant magnetic field (6) demonstrate that three different domains occur in the (H, T) phase diagram (figure 3), drawn here by the contour line $T_\alpha$ (H) where $\alpha$ reaches its optimum. The dashed areas visualize the low temperature regime with the weak (Pa) and highly polarized phase (PP) below and above $H_M$. These features were confirmed by extensive magnetization, specific heat and transport measurements (24-28).

At H = 0, analysis of the neutron scattering, specific heat and transport data is satisfactory in the frame of the so called self consistent renormalized spin fluctuation model (29-30). Extending this treatment to the magnetic field neutron scattering results gives the field variation of the ferromagnetic J(0) and antiferromagnetic J(k) exchange coupling (figure 4) (30-31); that corresponds to the strong field collapse of the dynamical antiferromagnetic correlation observed here at 1.6 meV (figure 5). A static ferromagnetic component has only been observed (Bragg Peak) (figure 5) ; in the range 0.4 – 10 meV no magnetic excitation has been observed. A strong dynamical independent q wavevector component persist though $H_M$.

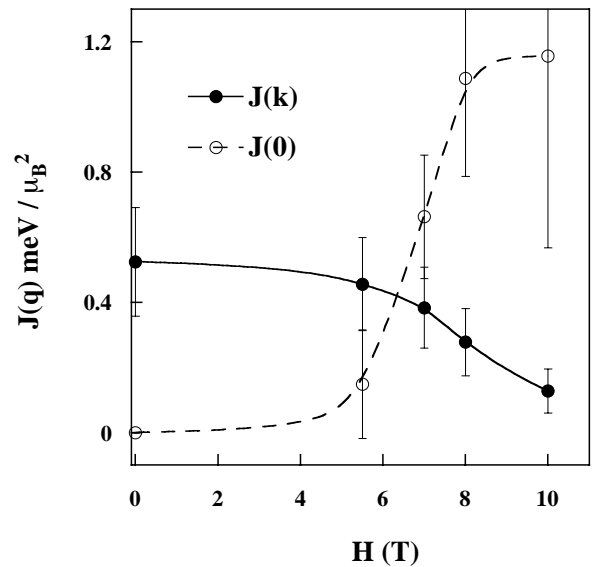

Figure 4 : The field dependence of the exchange coupling ferromagnetic J(0) and antiferromagnetic J (k) derived from a crude spin fluctuation model (31)

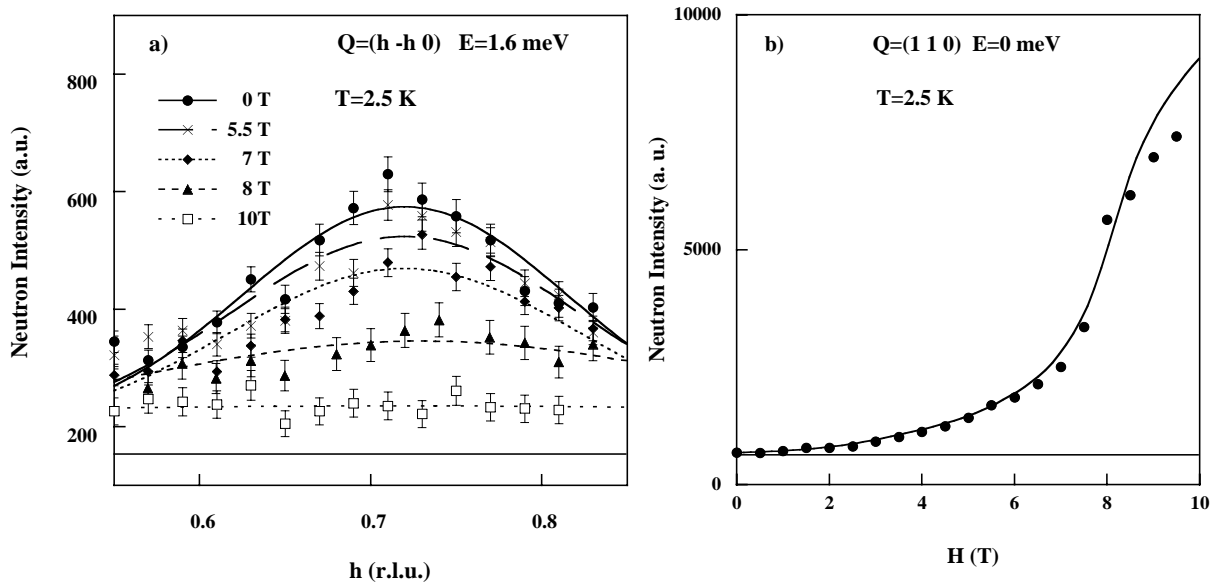

Figure 5 : Field results of $CeRu_2Si_2$ ; a) Q scans performed at 2.5 K at an energy transfert E = 1.6 mev, b) evolution of the maximum intensity at Q = (1,1,0) a ferromagnetic Bragg reflection at 2.5 K ; the line represents the square of the rescaled magnetization measured at 1.5 K (31)

Deep studies were also realized by dHVA experiments (22-23-32). Quantitatively below $H_M$, the results are well understood in the frame where the 4f electron is treated as itinerant (33); for the rather light electron or hole orbits complex behaviors are observed however only weak spin splitting occurs while differences seem to exist between the effective mass of spin up and spin down quasi particles. For the main large orbit $\psi$ detected below $H_M$, the dHVA signal is only observed for H closed to the basal plane (100) axis ; its effective mass m* reaches 120 $m_O$ (the free electron mass) (figure 6). Below $H_M$, the obtained FS lead to an electronic specific heat in excellent agreement with the experiments. Above $H_M$ there is no track of the $\psi$ orbit, a new hole orbit $\omega$ is detected now for H closed to the $|001)$ axis with rather moderate effective mass. Such trajectory is predicted in the band treatment where the f electron is assumed to be localized (34). However as now the measured FS is too small to explain the still large contribution of the electronic specific heat, parts of FS are missing. The physical idea is that the minority spin band gets a high effective mass despite the fact the occupation decreases (5).

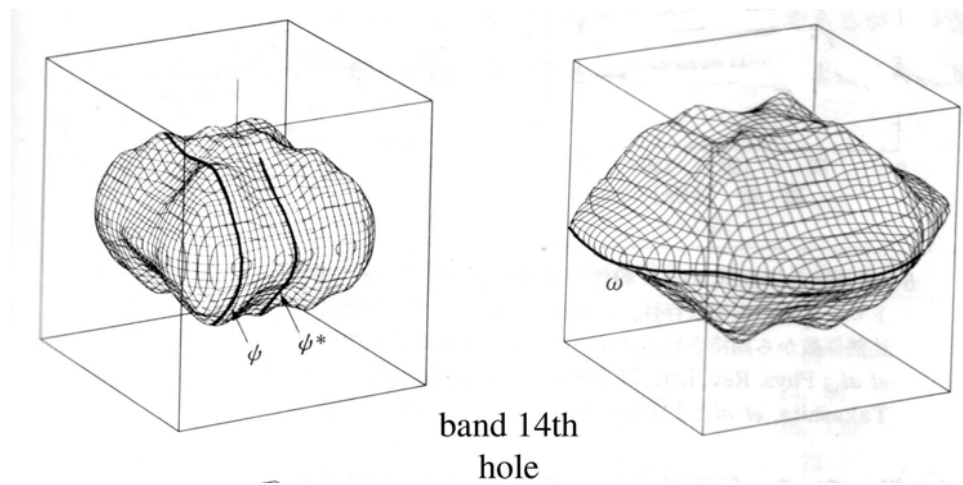

band 14th hole

Figure 6 : Hole Fermi surface observed, below $H_M$, $\Psi$ and, above $H_M$, $\omega$. See reference (22) for a complet description and discussion of the date.

If $T_K = 0$, the electronic low energy excitations is those of the host metal $LaRu_2Si_2$ ; the RKKY magnetic oscillations of these light electrons will give the indirect magnetic interactions between the quenched 4f electrons. For $T_K$ finite and at low temperature $T < T_K$ as pointed out qualitatively (35) and more quantitatively through for example the treatment of the so called exhaustion counting principle (36), on doping $LaRu_2Si_2$ with finite Ce atoms, drastic effects must appear in the spin and charge dynamics by comparison to the single site Kondo behavior. In the Anderson lattice, the 4f electron is itinerant whatever is the strength of the Coulomb repulsion U ; increasing U will modify the height of the Fermi discontinuity $Z \sim m^{*-1}$ of the distribution at the Fermi level (37).

In the Kondo lattice view, when the Kondo occurs via the many body interaction between 4f electrons and conduction electrons, new bands are formed where the both electrons are hybridized. This new FS is larger than the previous one based on the 4f localized model. Figure 6 illustrates the shrinking of the hole band between localized and itinerant descriptions. Theoretical discussions can be found in the references (38-41). Above $T_K$, studies by photoemission spectroscopy (42) indicates the exclusion of the 4f electron from $CeRu_2Si_2$ FS in good agreement with references (38-39).

In the crossover transition at $H_M$ no sudden change of FS will appear. By continuous increase of H, one will reach a fully polarized phase of $CeRu_2Si_2$ with a 4f level deep inside the conduction band and thus a wipe out of the Kondo structure at the Fermi level. It is striking that parts of FS reflects already above $H_M$ the topology of LaRu$_2$Si$_2$ FS. The nesting of the spin density wave **k** must be governed by the CeRu$_2$Si$_2$ topology of the FS at P = 0 ; it is also connected with the electronic anisotropy of the bare lattice of LaRu$_2$Si$_2$. For simplicity, only the AF correlations at **k** = **k$_1$** = (0.31, 0, 0) have been discussed ; two other AF correlations have been found for **k$_2$** = (0.31, 0.31, 0) and **k$_3$** = (0, 0, 0.35) (43).

Another important observation in CeRu$_2$Si$_2$ is that the metamagnetism appears under pressure for a critical value of the magnetization. That leads to suggest that the entropy can be expressed in the form $S = S\left(\dfrac{T}{T_S(P)}, \dfrac{H}{H_S(P)}\right)$ (6). Different theoretical proposals have been given ; so far the physics seems well described (with a pseudogap structure in the density of states of quasiparticles) either by a Hubbard model (44) or by periodic Anderson model (45).

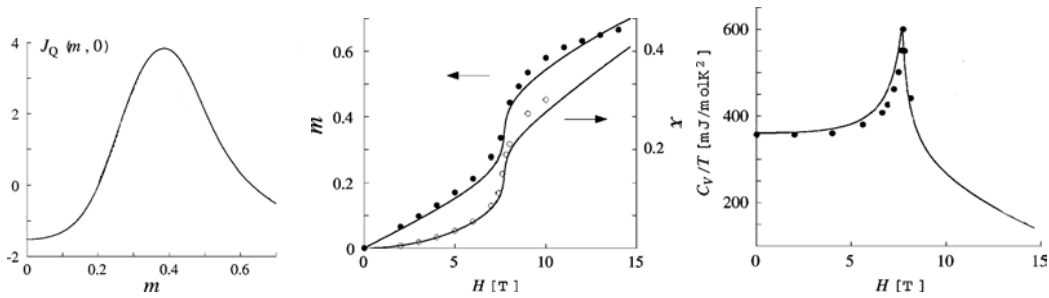

Figure 7 : Evolution of the exchange interaction J$_Q$ (M, 0) as a function of the magnetization M (in units of $k_B T_K$) and field dependences of the magnetization, the magnetostriction and C/T at constant volume at T = 0K (45)

In a lattice due to the multiplicity of the conduction bands a pseudogap structure is expected in the density of states ; in the recent calculation of the Anderson lattice (45), it was introduced directly. Since the magnetic exchange interaction J(Q) caused by virtual exchange of pair excitations of quasiparticles depends on the structure of the density of states, the field sweep will produce a drastic feedback on J(Q) with a change sign at $H_M$ as shown figure 7. The volume dependence of the Kondo temperature $T_K$ (x) is introduced simply through the enhancement produced by the large electronic parameter $\Gamma$ :

$$T_K(x) = T_K(0)\, e^{-x} \quad \text{with } x = \Gamma_e \dfrac{\Delta V}{V_0}$$

As the volume (V) dependence of J$_Q$ (M, x) is the same that $T_K$ i.e J$_Q$ (M,o)e$^{-x}$ ; J$_Q$ (M, x) will scaled with $T_K$. That explains the previous observed single parameter scaling ; the figure (7) shows the derived magnetization and magnetostriction at low temperature.

Let us point out that even for H = 0, low temperature (T < 1K) must be achieved for entering in a regime with a independent temperature Grüneisen parameter $\Gamma_e = -\partial \text{Log} T_S / \partial \text{Log} V$ which is caracteristic of an entropy law in S(T/T$_S$) (11). As for normal metals, these considerations are justified only for T < $\dfrac{E_F}{10 k_B}$ ie here $\dfrac{T_K}{10}$ with $T_K$ ~20 K. The remarkable fact is also the quasi equality between the zero field electronic Grüneisen parameter $\Gamma_e$ and the field Grüneisen parameter $\Gamma_{H_M} = -\dfrac{\partial Log H_M}{\partial Log V}$ (11-13).

Comparison with the bilayer ruthenate $Sr_3Ru_2O_7$

In the case of $CeRu_2Si_2$, the Ce spin has a strong Ising character and the Fermi surface has no specific low dimensionnal character. In the case of $Sr_3Ru_2O_7$, the Ru effective spin has mainly a Heisenberg character but 2 dimensional electronic character appear as in the single layer of $Sr_2RuO_4$. Quantitatively, the (H, T) "crossover" phase diagram reported is rather similar to that derived for $CeRu_2Si_2$ (7-46). The macroscopic common features are the pronounced temperature maxima of the initial susceptibility near 20K, the magnetization curves with a marked inflection point on cooling, the field dependence of the specific and associated effects in magnetoresistivity. Let us also notice that at $H_M$, the two compounds show a $\gamma$ enhancement near 60%.

Furthermore recent inelastic neutron scattering experiments show in low field two dimensionnal incommensurate magnetic fluctuations and the absence of long range antiferromagnetic ordering at least down to 1.5 K. As observed in $CeRu_2Si_2$, these correlation disappears on warming but here they become predominately ferromagnetic above 20 K (47).

For $CeRu_2Si_2$ no ferromagnetic fluctuation can be detected on warming when the AF correlation collapses (also associated with a sign change of the magnetoresistance). Our view is that $Sr_3Ru_2O_7$ in magnetic field will have a behavior rather similar of $CeRu_2Si_2$ with a field change of the magnetic exchange at $H_M$. As underlined there is of course differences in the spin space, electronic dimensionality and weight of the local fluctuations (large for heavy fermion compound like $CeRu_2Si_2$). Qualitatively the main difference between $CeRu_2Si_2$ and $Sr_3Ru_2O_7$ seems to appear in the inelastic contribution of the resistivity in the vicinity of $H_M$.

In $CeRu_2Si_2$ the Fermi liquid contribution in $T^2$ appear very robust, the main effects is the enhancement of the amplitude $A_H$ terme of the $A_H T^2$ law and the large increase of the residual resistivity. The range $T_I(H)$ of the very low temperature Fermi liquid $A_H T^2$ is weakly field depend and far to reproduce the $T_\alpha(H)$ phase diagram. Comparison of the data ($\rho_{exp}$) with the theoretical fit ($\rho_{th}$) obtained from the spin fluctuation parameters derived from the temperature dependence of the specific heat and from the dynamical susceptibility $\chi(q, \omega)$ results shows that an extra source of an electronic conduction is needed since $\rho_{data} < \rho_{th}$ (48). One possibility is the decoupling between hot and cold spots (49) ; in $CeRu_2Si_2$, the different parts of the FS are channels for parallel contributions and thus to differences between $T_I(H)$ and $T_\alpha(H)$.

In $Sr_3Ru_2O_7$, a $T^3$ contribution seems to emerge for H closed to $H_M$. The $T^3$ dependence of $\rho$ may mask the emergence of a supplementary contribution due to the crossing of an optimal in the ferromagnetic fluctuation. Applying a magnetic field may also push the carrier in a regime where the limitation is no more given by the collision between quasiparticules but by the FS topology and related changes. This puzzling result is still unexplained and need for clarification careful neutron scattering measurements and magnetorestriction studies through $H_M$.

## Polarized state of $CeRu_2Si_2$ versus ferromagnetism of $CeRu_2Ge_2$
## Competition between AF and F phase

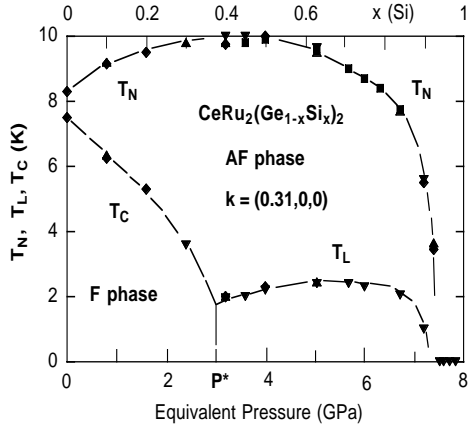

Figure 8 : (T, P) phase diagram of $CeRu_2Ge_2$ and of $CeRu_2(Ge_{1-x}Si_x)_2$ with the successive cascade of ferromagnetic ground state ($T_C$) up to $P^* \sim 3$ GPa, AF ($T_L$) and P above $P_C = 8.5$ GPa. $T_N$ is the Pa AF transition line ; the system changes slightly its AF ordering below $T_L$ (squaring) (50-53 and 55)

   $CeRu_2Ge_2$ realizes the situation where the lattice of $CeRu_2Si_2$ will be expanded by a virtual negative pressure near 8 GPa . For a positive pressure of 8 GPa the electronic properties of $CeRu_2Ge_2$ reproduces the same behavior than $CeRu_2Si_2$ at P = 0. As shown in figure (8) in the (T, P) phase diagram, at P = 0, $CeRu_2Ge_2$ presents two successive magnetic states on cooling . The first AF phase has the same incommensurate propagation vector $k_1$ than that found previously while the low temperature phase is ferromagnetic. This last phase disappears rapidly under pressure at P = P*. Above 3 GPa, only the AF order survives (50-53); the temperature $T_L$ describes the line where the modulated structure squares partly. FS measurements show that at P = 0, the f electron appears localized ie the orbits are those found in band calculation for $LaRu_2Si_2$ (54).

   Thus the transition of F to AF ground states at P* appears discontinuous. The simple picture is that it coincide with a discontinuous change in FS by contrast with the previous case of $CeRu_2Si_2$ through $H_M$ at P = 0. Recently the transition from ferromagnetic (F) to AF ground state (55) has been precised by a fine P study of the alloy $CeRu_2(Ge_{0.7}Si_{0.3})_2$ which realizes at P = 0 the situation of $CeRu_2Ge_2$ near 3 GPa. The Curie temperature $T_C$ does not collapse but collides at a finite temperature of 1.6 K with $T_L$ for P = 6 kbar. A quasivertical first order transition line is expected with discontinuities in entropy and volume according to the Clapeyron relation.

   In order to summarize this discussion, we have presented figure 9 the magnetic specific heat data of different compounds of $CeRu_2Si_2$ or $CeRu_2Ge_2$ families at P = 0. For $CeRu_2Ge_2$ , the extrapolated residual value of C/T term of the AF phase seems to reach an amplitude $\gamma$ very near to the critical value $\gamma_C = 600$ mJ mole$^{-1}$ K$^{-2}$ found for the critical concentration $x_C = 0.08$ in La substituted alloys by lanthanum and for AF ordered state x = 0.13 before it ends up in another phase below $T_L$. On the paramagnetic side ($x < x_C$), one can observe the slow continuous increase of C/T on cooling (x = 0.075 < $x_C$) with the difficulty to reach the low temperature Fermi liquid regime (8-10) ($T < T_I$) as $T_I \to 0$ at QCP. The large temperature plateau observed for $x = 0.013 > x_C$ seems to reproduce the balance of an increase of C/T on cooling due to the proximity of the magnetic instability and a decrease due to the entrance in an ordered magnetic phase (see $CeRu_2Ge_2$ steep decrease below $T_C$) ; obviously the longitudinal fluctuations appear decoupled from the static staggered magnetization.

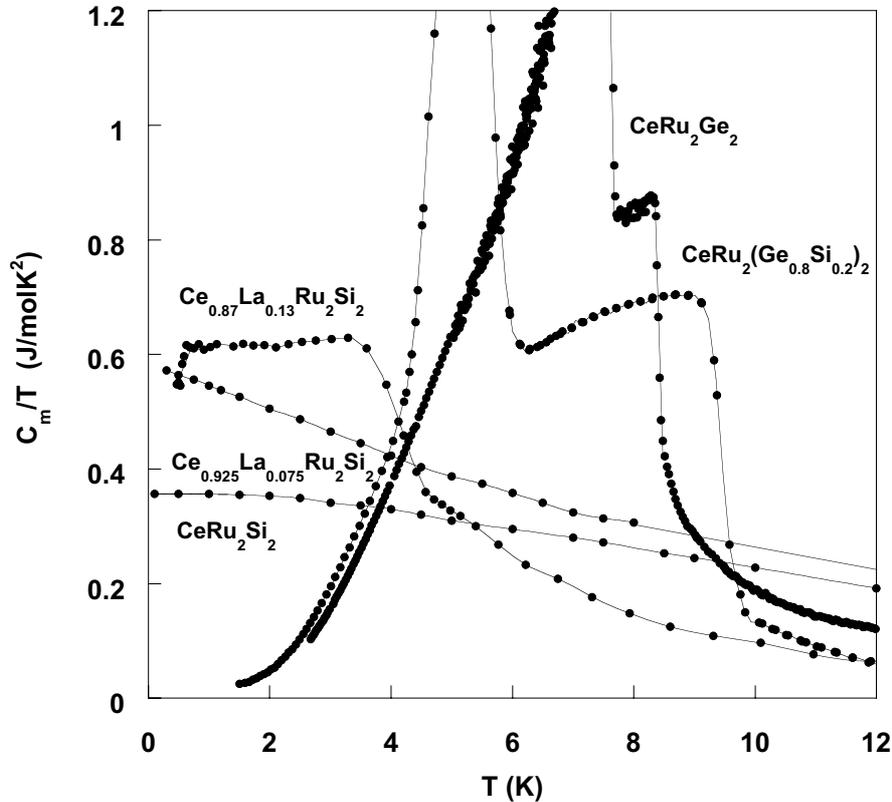

Figure 9 : Magnetic specific heat (ref. 8-10) divided by T versus for $CeRu_2Si_2$ $Ce_{0.925}La_{0.075}Ru_2Si_2$ $Ce_{0.87}La_{0.13}Ru_2Si_2$, $CeRu_2(Ge_{0.8}Si_{0.2})$ and $CeRu_2Ge_2$. lines are guides for the eyes.

Similar plateaus of C/T on the AF side are observed for other heavy fermion ordered state closed to $x_C$ or $P_C$ : for example for $Ce(Pd_{1-x}Ni_x)_2Ge_2$ ($x_c = .95$) (56) for $CeCu_{6-x}Au_x$ ($x_c=0.1$) (57) and for $CeCu_{6-x}Ag_x$ ($x = 0.2$) (58). In the case of the $CeRu_2Si_2$ serie starting from the paramagnetic side, the residual $\gamma$ = C/T term reaches its maxima value at $x_c$. This value seems to be preserved in the ordered phase closed to $x_c$ or $P_C$. However, below $T_L$ in the low temperature AF phase, a drop appears in $\gamma$ (see $Ce_{0.87}La_{0.13}Ru_2Si_2$). In agreement with the spin fluctuation prediction, the maxima of $\gamma$ will be at $x_c$ ; the theory predicts a $\sqrt{x - x_c}$ or $\sqrt{P - P_C}$ dependence (12).

In figure (9), for $Ce_{0.87}La_{0.13}Ru_2Si_2$ first below $T_N$, two different components seem to coexist static and dynamic which correspond respectively to the establishment of the sublattice magnetization and to a freezing to the low energy excitation to the QCP down to $T_L$ (C/T = $\gamma_c$ $T_N > T > T_L$). Pressure inelastic neutron scattering experiments on the same alloy confirm the quasi invariance of the spin dynamic through $P_C$ (59). In the $CeCu_6$ serie, at least down to 40 mK, the maxima extrapolated value of $\gamma$ occurs in the AF domain (57) ; this striking result escapes from an usual spin fluctuation prediction. Possible arguments for this deviation are the incomplet formation of the quasiparticle band (as its bandwidth is now extremely small T ~5K) and also the rather large disorder introduced by the substitution of Cu sites (30).

For $CeRu_2Ge_2$, the entropy conservation precludes a large C/T term down to 0K characteristic of the AF phase ; through a first order transition, a large drop of entropy will occur at $T_C$ with the concomitant drop of $\gamma$ from $\gamma \sim \gamma_C$ for the AF phase to $\gamma \sim 16.5$ mJ mole$^{-1}$ K$^{-2}$ for the F phase. This drastic modification is associated with the related deep modification of the FS. By analogy to the previous consideration, the previous pseudogap periodic Anderson model scheme (45) gives a good qualitative idea why for a high value of the sublattice magnetization $M_0$ ferromagnetic ground state is favored, and why, for a low critical value of the sublattice magnetization $M_0$, AF will win.

Finally let us point out the difference in (H, T) phase diagram between AF and Pa phase in the $Ce_{1-x}La_xRu_2Si_2$ serie. For x = 0.1, 0.13 (at P = 0) and x = 0.2 at different pressures, it was clearly observed two transition fields $H_a$ and $H_c$ quasi invariant in P on approaching the QCP (x → $x_c$, P → $P_C$) from the AF side with a supplementary evidence of the pseudo-metamagnetic field $H_M$ strongly pressure dependent ($\Gamma_{H_M}$). At the QCP, ($T_N = 0$), $H_M$ collides at T = 0K with $H_c$ the true critical field defining the transition between the AF and Pa phases ; when $T_N$ (H = 0) increases, the intercept of $H_c$ and $H_M$ occurs at finite temperature (10-60).

Absence of unconventional superconductivity

$CeRu_2Si_2$ and $CeRu_2Ge_2$ with respective dominant AF and F interactions are excellent candidate to understand the occurrence of superconductivity by spin fluctuation mechanisms. In a simple picture, one can expect unconventional superconductivity with respective singlet and triplet pairing (61).
In $CeRu_2Si_2$ at least down to 20 mK, no track of superconductivity has been found. By contrast, superconductivity has been discovered when AF heavy fermion systems like $CeCu_2Ge_2$ (62), $CePd_2Si_2$ (63) and $CeIn_3$ (64-65) are tuned through their magnetic QCP. The microscopic reasons of a persistent normal phase can be the weakness of a pairing through AF spin fluctuation caused by the lack of a transverse component (Ising character of the magnetism) (66) and/or the sharpness of the superconducting domain centered at $P_C$ (67). If the superconducting temperature $T_S$ is low, the necessary clean condition for the occurrence of unconventional superconductivity $\varphi_S < l_e$ ie superconducting coherence lengh ($\varphi$) smaller than the electronic path $l_e$) may be quite restrictive ($\varphi_S \sim T_S^{-1}$).

One can also argue why superconductivity is not observed in the ferromagnetic state of $CeRu_2Ge_2$ near P* at the opposite cases of $UGe_2$ (68) and $URhGe$ (69). No systematic experiments have been yet realized. However even just below P*, the attractive spin fluctuation potential may be weak for a triplet pairing as m* has still a low P value (50).

Conclusion

We have underlined the main results of the $CeRu_2Si_2$ heavy fermion lattice with special focus on its so called metamagnetism. The comparison of the magnetic field sweep in $CeRu_2Si_2$ and pressure sweep in $CeRu_2Ge_2$ confirms the trends found in the field neutron scattering studies of $CeRu_2Si_2$ with the interplay between the 4f itinerary, the band structure, the nature of the exchange coupling and the lattice instabilities. The critical behavior found at $H_M$ in $CeRu_2Si_2$ is rather reminiscent of the so called Lifshitz point case (70). We also stress that the recent discovered metamagnetism of $Sr_3Ru_2O_7$ belongs to the same class of phenomena ; fair comparison with $CeRu_2Si_2$ will be worthwhile. Excellent introduction and discussion of heavy fermions can be found in the textbook of reference 71.

One of the author, JF, thanks Pr. A. McKenzie for his indirect stimulation in comparing $CeRu_2Si_2$ with $Sr_3Ru_2O_7$ . That pushes us to precise the case of $CeRu_2Si_2$ . We use the opportunity to thank Pr. J. Franse for his continuous interest as well as for the efficiency of his collaborator A. de Visser in experiments performed between Amsterdam and Grenoble a decade ago.